\documentclass[12pt]{article}
\usepackage{stmaryrd}
\usepackage{bbm}
\usepackage{mathrsfs}
\usepackage{amssymb}
\usepackage{amsmath}
\usepackage{amsthm}
\usepackage{graphicx}
\usepackage{color}
\usepackage{ulem}
\usepackage[square,numbers,sort&compress]{natbib}

\newtheorem{theorem}{Theorem}

\newtheorem{lemma}{Lemma}

\allowdisplaybreaks[4]
\begin{document}

\title{On realizations of the Witt algebra in $\mathbb{R}^3$\thanks{Supported by
the NSF of China (Grant Nos. 11101332, 11201371, 11371293), the Foundation of
Shaanxi Educational Committee, China (Grant No. 11JK0482) and the
NSF of Shaanxi Province, China (Grant No. 2012JQ1013). } }

\author{{ Renat Zhdanov$^{1}$ and Qing Huang$^{2,3}$}\\
{\small 1. BIO-key International, 55121 Eagan, MN, USA}\\
{\small 2. Department of Mathematics, Northwest University, Xi'an 710069, China}\\
{\small 3. Center for Nonlinear Studies, Northwest University, Xi'an 710069, China}}
\date{}

\maketitle
\begin{abstract}

We obtain exhaustive classification of inequivalent
realizations of the Witt and Virasoro algebras by Lie vector
fields of differential operators in the space $\mathbb{R}^3$.
Using this classification we describe all inequivalent
realizations of the direct sum of the Witt algebras in $\mathbb{R}^3$.
These results enable constructing all possible (1+1)-dimensional classically
integrable equations that admit infinite dimensional symmetry algebra
isomorphic to the Witt or the direct sum of Witt algebras. In this way
the new classically integrable nonlinear PDE in one spatial dimension has
been obtained. In addition, we construct a number of new nonlinear
(1+1)-dimensional PDEs admitting infinite symmetries.
\end{abstract}

\section{Introduction}

Since its introduction in the 19th century, Lie group analysis has
become a very popular and powerful tool for solving nonlinear partial
differential equations (PDEs). Given a PDE that possesses a nontrivial
Lie symmetry, we can utilize symmetry reduction procedure to construct
its exact solutions \cite{fus93,zhd97}.

Not surprisingly, the wider symmetry of an equation under study
is, the better off we are when applying the Lie approach to solve it. This is
especially the case when its symmetry group is infinite-parameter. If a
nonlinear differential equation admits infinite Lie symmetries, then it
is often possible either to linearize it or construct its general solution
\cite{fus93}.

The classical example is the hyperbolic type Liouville equation
\begin{equation}
\label{lio}
u_{tx}=\exp(u),
\end{equation}
which admits the infinite-parameter Lie group
\begin{equation}
\label{inf}
t'=t+f(t),\quad x'=x+g(x),\quad u'=u-\dot f(t)-\dot g(x),
\end{equation}
where $f$ and $g$ are arbitrary smooth functions. The general solution
of Eq. (\ref{lio}) can be obtained by the action of transformation group
(\ref{inf}) on its particular traveling wave solution
of the form $u(t,x)=\varphi(x+t)$ (see, e.g., \cite{fus93}). An alternative
way to solve the Liouville equation is linearization \cite{fus93}.

Note that the Lie algebra of Lie group (\ref{inf}) is the direct
sum of two infinite-dimensional Witt algebras, which are subalgebras
of the Virasoro algebra.

Unlike the finite-dimensional algebras, infinite-dimensional
ones have not been systematically studied within the
context of classical Lie group analysis of nonlinear PDEs.
The situation is, however, drastically different in the
case of generalized (higher) Lie symmetries which played the critical
role in success of the theory of integrable systems in $(1+1)$- and
$(1+2)$-dimensions (see, e.g. \cite{ibr85}).

The breakthrough in the analysis of integrable systems has been nicely
complemented by development of the theory of infinite-dimensional
Lie algebras such as loop \cite{pre86}, Kac-Moody \cite{kac94} and
Virasoro algebras \cite{ioh11}.

Virasoro algebra plays an increasingly important role in
mathematical physics in general \cite{bel84, gra13} and in the theory
of integrable systems in particular. Study of nonlinear
evolution equations in (1+2)-dimensions arising in different
areas of modern physics shows that many of these equations admit
Virasoro algebras as their symmetry algebras. Let us mention among others the
Kadomtsev-Petvishilvi (KP) \cite{dav85, dav86,gun02}, modified KP,
cylindrical KP \cite{levi88}, the Davey-Stewartson \cite{cha88, gun06},
Nizhnik-Novikov-Veselov, stimulated Raman scattering, (1+2)-dimensional
Sine-Gordon \cite{sen98} and the KP hierarchy \cite{orl97} equations.

It is a common belief that nonlinear PDEs admitting symmetry algebras
of Virasoro type are prime candidates for the roles of integrable systems.
Consequently, systematic classification of inequivalent realizations of
the Virasoro algebra is a crucial step of symmetry approach to
constructing integrable systems (see, e.g., \cite{lou95,lou04}).
It should be pointed out that there are a few integrable
equations which do not possess Virasoro symmetry algebras, such as the
breaking soliton and Zakharov-Strachan equations \cite{sen98}.

Classification of Lie algebras of vector fields of differential
operators within the action of local diffeomorphism group has been
pioneered by Sophus Lie himself. It remains a very powerful method
for group analysis of nonlinear differential equations.
Some of the more recent applications of this approach include geometric
control theory \cite{jur97}, theory of systems of nonlinear
ordinary differential equations possessing superposition principle
\cite{shn84}, algebraic approach to molecular dynamics \cite{alh84, sal10} to
mention only a few. Still the biggest bulk of results has been
obtained in the area of classification of nonlinear PDEs possessing
point and higher Lie symmetries (see \cite{zhd01} and references
therein). Analysis of realizations of Lie algebras by first-order
differential operators is in the core of almost every approach to
group classification of PDEs (see, e.g.,
\cite{ack75, cam66,gon92a,kam89,kos83,gon92,zhd97})

In this paper we concentrate on the realizations of
the Witt and Virasoro algebras by first-order differential operators
in the space $\mathbb{R}^n$ with $n\le 3$. One of our primary
motivations was that with these realizations in hand we can develop
a regular way to construct $(1+1)$-dimensional nonlinear PDEs
which are integrable in the sense that they admit infinite symmetries.

The paper is organized as follows. In Section 2 we give a brief
account of necessary facts and definitions. In addition, the
algorithmic procedure for realizations of the Virasoro algebra is
described in detail. We construct all inequivalent realizations of
the Witt algebra (a.k.a. centerless Virasoro algebra) in Section 3.
Section 4 is devoted to the description of the realizations of
the Virasoro algebra. We prove that there are no central extensions
of the Witt algebra in the space $\mathbb{R}^3$. In Section 5 we
construct broad classes of nonlinear PDEs admitting infinite dimensional
symmetry algebras, which are realizations of the Witt algebra. Furthermore,
all inequivalent realizations of the direct sum of two Witt algebras
are obtained in Section 6. This enables us to classify the second-order PDEs whose invariance algebra
contains a direct sum of the Witt algebras. We prove that any such
PDE is equivalent to one of the four canonical equations (\ref{d1})--(\ref{d6}).
The last section contains a brief summary of the obtained
results.

\section{Notations and definitions}

The Virasoro algebra, $\mathfrak{V}$, is the infinite-dimensional Lie
algebra with basis elements $\{L_n,\ n\in\mathbb{Z}\}\bigcup \{C\}$
which satisfy the commutation relations
\begin{equation*}
[L_m,\ L_n] = (m-n)
L_{m+n}+\frac{1}{12}m(m^2-1)\delta_{m,-n}C,\quad [L_m,C]=0,\quad
m,n\in\mathbb{Z},
\end{equation*}
where $[Q, P]=QP-PQ$ is the commutator of Lie vector fields $P$ and
$Q$, and $\delta_{a,b}$ stands for the Kronecker delta
\begin{equation*}
\delta_{a,b}=\left\{\begin{array}{ll}1, &a=b,\\[2mm]
    0, & {\rm otherwise}.\end{array}\right.
\end{equation*}
The operator $C$ commuting with all other basis elements is called the
central element. If $C$ equals to zero, the algebra $\mathfrak{V}$
reduces to the centerless Virasoro algebra or Witt algebra $\mathfrak{W}$.
Consequently, the full Virasoro algebra is the nontrivial one-dimensional
central extension of the Witt algebra.

We now consider the Virasoro algebras as the linear subspace of the
infinite-dimensional Lie algebra $\mathfrak{L}_{\infty}$ spanned by
the basis elements of the form
\begin{equation}\label{basis}
Q=\tau(t,x,u)\partial_t+\xi(t,x,u)\partial_x+\eta(t,x,u)\partial_u
\end{equation}
over $\mathbb{R}^3$. Applying the transformation
\begin{equation}\label{tr}
t\to \tilde{t}=T(t,x,u),\quad x\to \tilde{x}=X(t,x,u),\quad u\to
\tilde{u}=U(t,x,u),
\end{equation}
with $D(T,X,U)/D(t,x,u)\neq0$ to \eqref{basis}, we get
\begin{equation*}
\tilde{Q}=(\tau T_t+\xi T_x+\eta T_u)\partial_{\tilde{t}}+(\tau
X_t+\xi X_x + \eta X_u)\partial_{\tilde{x}}+(\tau U_t+\xi U_x+\eta
U_u)\partial_{\tilde{u}}.
\end{equation*}
Evidently, $\tilde{Q}\in\mathfrak{L}_{\infty}$. Hence we see that
the set of operators \eqref{basis} is invariant with respect to the
transformation \eqref{tr}.

It is well-known that the correspondence, $Q$ $\sim$ $\tilde{Q}$, is the
equivalence relation and as such it splits the set of operators
(\ref{basis}) into some equivalence classes. Any two elements within the same
equivalence class are related through a transformation \eqref{tr},
while two elements belonging to different classes
cannot be transformed one into another by a transformation of the
form \eqref{tr}. Hence to describe all possible realizations of the
Virasoro algebra, one needs to construct a representative of each
equivalence class. The remaining realizations can be obtained by applying
transformations (\ref{tr}) to the representatives in question.

To construct all inequivalent realizations of the Virasoro
algebra we need to implement the following steps:
\begin{itemize}
\item Describe all inequivalent forms of $L_0$, $L_1$ and $L_{-1}$
such that the commutation relations of the Virasoro subalgebra,
\begin{equation} \label{sub}
[L_0,L_1]=-L_1,\quad
[L_0,L_{-1}]=L_{-1},\quad [L_1,L_{-1}]=2L_0,
\end{equation}
hold together with the relations $[L_i,C]=0,\ (i=0,1,-1)$.
Note that algebra $\langle L_0,L_1,L_{-1}\rangle$ is isomorphic to $sl(2,\mathbb{R})$.
\item Construct all inequivalent realizations of the operators $L_2$
and $L_{-2}$ which commute with $C$ and satisfy the relations:
\begin{equation}\label{5d}
\begin{array}{c}
[L_0,L_2]=-2L_2,\quad [L_{-1},L_2]=-3L_1,\quad [L_1,L_{-2}]=3L_{-1},\\[2mm]
[L_0, L_{-2}]=2L_{-2},\quad[L_2,L_{-2}]=4L_0+\frac12C.
\end{array}
\end{equation}
\item Derive all the remaining basis operators of the Virasoro algebra
through the recursion relations
\begin{equation*}
L_{n+1}=(1-n)^{-1}[L_1,\ L_n],\quad L_{-n-1}=(n-1)^{-1}[L_{-1},\
L_{-n}]
\end{equation*}
with
\begin{equation*}
   [L_{n+1},L_{-n-1}]=2(n+1)L_0+\frac{1}{12}n(n+1)(n+2)C,\quad [L_i,C]=0,
\end{equation*}
where $i=n+1,-n-1$ and $n=2,3,4,\cdots$.
\end{itemize}

In Sections 3 and 4, we will implement the algorithm above to construct
all inequivalent realizations of the Witt and Virasoro
algebras by operators (\ref{basis}).

\section{Realizations of the Witt algebra}
Turn now to describing realizations of the Witt algebra $\mathfrak{W}$.
Let us remind that the algebra $\mathfrak{W}$ is obtained from the Virasoro
algebra by putting $C=0$. We begin by letting the vector field $L_0$ be of
the general form \eqref{basis}, namely,
\begin{equation*}
L_0=\tau(t,x,u)\partial_t+\xi(t,x,u)\partial_x+\eta(t,x,u)\partial_u.
\end{equation*}
Transformation \eqref{tr} maps $L_0$ into
\begin{equation*}
\tilde{L}_0=(\tau T_t+\xi T_x+\eta T_u)\partial_{\tilde{t}}+(\tau
X_t+\xi X_x+\eta X_u)\partial_{\tilde{x}}+(\tau U_t+\xi U_x+\eta
U_u)\partial_{\tilde{u}}.
\end{equation*}
We have $\tau^2+\xi^2+\eta^2\neq0$, otherwise $L_0$ is trivial.
Consequently, we can choose the solutions of equations
\begin{equation*}
\tau T_t+\xi T_x+\eta T_u=1,\quad\tau X_t+\xi X_x+\eta X_u=0,\quad
\tau U_t+\xi U_x+\eta U_u=0.
\end{equation*}
as $T$, $X$ and $U$ and reduce $L_0$ to the form $L_0=\partial_t$
(hereafter we drop the tildes). Then $L_0$ is
equivalent to the canonical operator $\partial_t$.

With $L_0$ in hand we now proceed to constructing $L_1$ and $L_{-1}$
which obey the commutation relations \eqref{sub}. Letting $L_1$ be of
the general form \eqref{basis} and inserting it into $[L_0,L_1]=-L_1$
yield
\begin{equation*}
L_1=\mathrm{e}^{-t}f(x,u)\partial_t+\mathrm{e}^{-t}g(x,u)\partial_x
+ \mathrm{e}^{-t}h(x,u)\partial_u,
\end{equation*}
where $f,\ g,\ h$ are arbitrary smooth functions. To further
simplify $L_1$ we use an equivalence transformation of the form
\eqref{tr} preserving $L_0$. Applying \eqref{tr} to $L_0$ gives
\begin{equation*}
L_0\to\tilde{L}_0=T_t\partial_{\tilde{t}}+X_t\partial_{\tilde{x}}
+U_t\partial_{\tilde{u}}=\partial_{\tilde{t}}.
\end{equation*}
Hence, transformation
\begin{equation*}
\tilde{t}=t+T(x,u),\ \tilde{x}=X(x,u),\ \tilde{u}=U(x,u)
\end{equation*}
is the most general transformation that does not alter the form of
$L_0$. It converts Lie vector field $L_1$ into
\begin{equation*}
\tilde{L_1}=\mathrm{e}^{-t}(f +gT_x+hT_u)\partial_{\tilde{t}}
+\mathrm{e}^{-t}(gX_x+hX_u)\partial_{\tilde{x}}
+\mathrm{e}^{-t}(gU_x+hU_u)\partial_{\tilde{u}}.
\end{equation*}
To further analyze the realizations of $L_1$, we need to
consider the inequivalent cases $g^2+h^2=0$ and $g^2+h^2\not=0$.

{\bf Case 1.} If $g^2+h^2=0$, we have
$\tilde{L_1}=\mathrm{e}^{-t}f(x,u) \partial_{\tilde{t}}$. Choosing
$\tilde{t}=t-\ln{|f(x,u)|}$ gives $L_1=\mathrm{e}^{-t}\partial_t$.
Let $L_{-1}$ be of the general form \eqref{basis}. And taking into account
\eqref{sub} we get $L_{-1}=\mathrm{e}^t\partial_t$.

{\bf Case 2.} Provided $g^2+h^2\neq0$, we choose $\tilde{t}=t+T(x,u)$,
where $T(x,u)$ satisfies the relation
\begin{equation*}
\mathrm{e}^{-T}=f+gT_x+hT_u,
\end{equation*}
and take $X$ and $U$ to be solutions of the equations
\begin{equation*}
gX_x+hX_u=\mathrm{e}^{-T},\quad gU_x+hU_u=0.
\end{equation*}
Then $L_1$ is mapped into $\mathrm{e}^{-t}(\partial_t+\partial_x)$.
Selecting $L_{-1}$ of the form \eqref{basis} and taking into
account commutation relations \eqref{sub}, we arrive at
\begin{equation*}
L_{-1}=\mathrm{e}^t(1-\mathrm{e}^{-2x}f_1(u))\partial_t
+\mathrm{e}^t(-1-\mathrm{e}^{-2x}f_1(u)
+\mathrm{e}^{-x}g_1(u))\partial_x
+\mathrm{e}^{t-x}h_1(u)\partial_u,
\end{equation*}
where $f_1,g_1,h_1$ are arbitrary smooth functions.

Acting by the transformation
\begin{equation}\label{tr2}
\tilde{t}=t,\quad \tilde{x}=x+X(u),\quad \tilde{u}=U(u),
\end{equation}
which keeps $L_0$ and $L_1$ invariant, on $L_{-1}$ gives
\begin{equation*}
\tilde{L}_{-1}=\mathrm{e}^t(1-\mathrm{e}^{-2x}f_1(u))\partial_{\tilde{t}}
+\mathrm{e}^t(-1-\mathrm{e}^{-2x}f_1(u)+\mathrm{e}^{-x}g_1(u)
+\mathrm{e}^{-x}h_1\dot{X})\partial_{\tilde{x}}
+\mathrm{e}^{t-x}h_1(u)\dot{U}\partial_{\tilde{u}}.
\end{equation*}

To complete the analysis, we consider the cases
$f_1(u)\neq0$ and $f_1(u)=0$ separately.

Assuming that $f_1(u)\neq0$ we choose
\begin{equation*}
X(u)=-\ln{\sqrt{|f_1(u)|}},\quad
\phi(u)=(g_1(u)+h_1(u)\dot{X}(u))/\sqrt{|f_1(u)|}.
\end{equation*}
In addition, we take as $U$ in \eqref{tr2} a solution of $h_1(u)\dot{U}=1$ if
$h_1\neq0$ or an arbitrary non-constant function if $h_1=0$.
This yields
\begin{equation*}
L_{-1}=\mathrm{e}^{t}(1+\alpha\mathrm{e}^{-2x})\partial_{t}
+\mathrm{e}^{t}(-1+\alpha\mathrm{e}^{-2x}+\mathrm{e}^{-x}\phi(u))\partial_{x}
+\beta\mathrm{e}^{t-x}\partial_{u},
\end{equation*}
where $\alpha=\pm 1$ and $\beta=0,1$.

The case $f_1(u)=0$ gives rise to the realization
\begin{equation*}
\tilde{L}_{-1}=\mathrm{e}^t\partial_{\tilde{t}}+\mathrm{e}^t(-1+
\mathrm{e}^{-x}g_1(u)+\mathrm{e}^{-x}h_1(u)\dot{X})\partial_{\tilde{x}}
+\mathrm{e}^{t-x}h_1(u)\dot{U}\partial_{\tilde{u}}.
\end{equation*}
Letting $X=0$ and $U$ in \eqref{tr2} be a solution of $h_1(u)\dot{U}=1$ when
$h_1\neq0$ or an arbitrary non-constant function otherwise, we
get
\begin{equation*}
L_{-1}=\mathrm{e}^t\partial_{t}+\mathrm{e}^t(-1+\mathrm{e}^{-x}g_1(u))\partial_{x}
+\beta\mathrm{e}^{t-x}\partial_{u}
\end{equation*}
with $\beta=0,1$.

\begin{lemma}\label{3d}
Any triplet of operators $\langle L_0,L_1,L_{-1}\rangle$ obeying the
commutation relations of the Witt algebra is equivalent to either
\begin{equation}\label{rep1}
\langle \partial_t,\ {\rm e}^{-t}\partial_t,\ {\rm e}^{t}\partial_t\rangle
\end{equation}
or
\begin{equation}\label{rep2}
\langle \partial_t,\ {\rm
e}^{-t}(\partial_t+\partial_x),\ {\rm e}^{t}(1+\alpha{\rm e}^{-2x})\partial_t
+{\rm e}^t(-1+\phi(u){\rm e}^{-x}+\alpha{\rm
e}^{-2x})\partial_x+\beta{\rm e}^{(t-x)}\partial_u\rangle
\end{equation}
Here $\alpha=0,\pm 1$, $\beta = 0,1$ and $\phi(u)$ is an arbitrary smooth function.
\end{lemma}

Now to obtain the complete description of all inequivalent Witt algebras
we need to extend algebras \eqref{rep1} and \eqref{rep2} by the
operators $L_{2}$ and $L_{-2}$ and implement the last two steps
of the classification procedure given in Section 2.

We first formulate the final results and then present the detailed proof.

\begin{theorem}\label{witt}
There are at most eleven inequivalent realizations of the Witt
algebra $\mathfrak{W}$ over the space $\mathbb{R}^3$. The representatives
$\mathfrak{W}_i$, $(i=1,2,\ldots,11)$ of each equivalence class
are listed below.
\begin{eqnarray*}
&\mathfrak{W}_1:&\langle\mathrm{e}^{-nt}\partial_t\rangle,\\[2mm]
&\mathfrak{W}_2:&\langle\mathrm{e}^{-nt}\partial_t+\mathrm{e}^{-nt}[n+\frac12n(n-1)\alpha\mathrm{e}^{-x}]\partial_x\rangle,\\[2mm]
&\mathfrak{W}_3:&\langle\mathrm{e}^{-nt+(n-1)x}[\mathrm{e}^{2x}-(n+1)\gamma\mathrm{e}^x+\frac12n(n+1)\gamma^2](\mathrm{e}^x-\gamma)^{-n-1}\partial_t\\[2mm]
               &&\qquad +\mathrm{e}^{-nt+(n-1)x}[n\mathrm{e}^x-\frac12n(n+1)\gamma](\mathrm{e}^x-\gamma)^{-n}\partial_x\rangle,\\[2mm]
&\mathfrak{W}_4:&L_0=\partial_t,\\[2mm]
               && L_1=\mathrm{e}^{-t}\partial_t+\mathrm{e}^{-t}\partial_x,\\[2mm]
               &&L_{-1}=\mathrm{e}^t(1+\gamma\mathrm{e}^{-2x})\partial_t+\mathrm{e}^t(-1+\gamma\mathrm{e}^{-2x}+\mathrm{e}^{-x}\tilde{\phi})\partial_x,\\[2mm]
               && L_2=\mathrm{e}^{-2t}f(x,u)\partial_t+\mathrm{e}^{-2t}g(x,u)\partial_x,\\[2mm]
               &&L_{-2}=\mathrm{e}^{2t}[1+3\gamma\mathrm{e}^{-2x}-\frac12{\mathrm{e}^{-3x}}(6\gamma\tilde{\phi}+\tilde{\phi}^3\pm(4\gamma+\tilde{\phi}^2)^{3/2})]\partial_t\\[2mm]
               &&\qquad+\mathrm{e}^{2t}[-2+3\mathrm{e}^{-x}\tilde{\phi}+6\gamma\mathrm{e}^{-2x}-\frac12{\mathrm{e}^{-3x}}
                 (6\gamma\tilde{\phi}+\tilde{\phi}^3\pm(4\gamma+\tilde{\phi}^2)^{3/2})]\partial_x,\\[2mm]
               && L_{n+1}=(1-n)^{-1}[L_1,\ L_n],\quad L_{-n-1}=(n-1)^{-1}[L_{-1},\ L_{-n}],\quad n\ge 2,\\[2mm]
&\mathfrak{W}_5:&\langle\mathrm{e}^{-nt+(n-1)x}(\mathrm{e}^{x}\pm n)(\mathrm{e}^x\pm1)^{-n}\partial_t+n\mathrm{e}^{-nt+(n-1)x}(\mathrm{e}^x\pm1)^{1-n}\partial_x\rangle,\\[2mm]
&\mathfrak{W}_6:&\langle\mathrm{e}^{-nt}\partial_t+\gamma\mathrm{e}^{-nt}[\mathrm{e}^{nx}-(\mathrm{e}^x-\gamma)^n](\mathrm{e}^x-\gamma)^{1-n}\partial_x\rangle,\\[2mm]
&\mathfrak{W}_7:& L_0=\partial_t,\\[2mm]
               && L_1=\mathrm{e}^{-t}\partial_t+\mathrm{e}^{-t}\partial_x,\\[2mm]
               &&L_{-1}=\mathrm{e}^t(1+\gamma\mathrm{e}^{-2x})\partial_t+\mathrm{e}^t(-1+\gamma\mathrm{e}^{-2x}+\mathrm{e}^{-x}\tilde{\phi})\partial_x,\\[2mm]
               && L_2=\mathrm{e}^{-2t+x}\frac{\mathrm{e}^x-\tilde{\phi}}{\mathrm{e}^{2x}-\mathrm{e}^x\tilde{\phi}-\gamma}\partial_t+
                \mathrm{e}^{-2t+x}\frac{2\mathrm{e}^x-\tilde{\phi}}{\mathrm{e}^{2x}-\mathrm{e}^x\tilde{\phi}-\gamma}\partial_x,\\[2mm]
               && L_{-2}=\mathrm{e}^{2t-3x}(\mathrm{e}^{3x}+3\gamma\mathrm{e}^x-\gamma\tilde{\phi})\partial_t+
                \mathrm{e}^{2t-3x}(2\mathrm{e}^x-\tilde{\phi})(-\mathrm{e}^{2x}+\mathrm{e}^x\tilde{\phi}+\gamma)\partial_x,\\[2mm]
               && L_{n+1}=(1-n)^{-1}[L_1,\ L_n],\quad L_{-n-1}=(n-1)^{-1}[L_{-1},\  L_{-n}],\quad n\ge 2,\\[2mm]
&\mathfrak{W}_8:&\langle\mathrm{e}^{-nt}\partial_t+\mathrm{e}^{-nt}[n-\mathrm{sgn}(n)\frac{\gamma}{2}\sum_{j=1}^{|n|-1}j(j+1)\mathrm{e}^{-2x}]\partial_x\rangle,\\[2mm]
&\mathfrak{W}_9:&\langle\frac{\mathrm{e}^{-nt+(n-1)x}}{(\mathrm{e}^x-1)^{n+2}}[(-1+\sum_{j=1}^{|n|-1}(2j+1))n+(2n+1)\mathrm{e}^x-(n+2)\mathrm{e}^{2x} +\mathrm{e}^{3x}\\[2mm]
            &&\qquad +\mathrm{sgn}(n)\frac{\tilde{\phi}}2\sum_{j=1}^{|n|-1}j(j+1)]\partial_t+\frac{\mathrm{e}^{-nt+(n-1)x}}{(\mathrm{e}^x-1)^{n+1}}
                 [(1-\sum_{j=1}^{|n|-1}(2j+1))n\\[2mm]
            &&\qquad -2n\mathrm{e}^x+n\mathrm{e}^{2x}
                  -\mathrm{sgn}(n)\frac{\tilde{\phi}}2\sum_{j=1}^{|n|-1}j(j+1)]\partial_x\rangle,\\[2mm]
&\mathfrak{W}_{10}:&\langle\mathrm{e}^{-nt}\partial_t+n\mathrm{e}^{-nt}\partial_x+
\frac{\mathrm{sgn}(n)}2\sum_{j=1}^{|n|}j(j-1)\mathrm{e}^{-nt-2x}\partial_u\rangle,\\[2mm]
&\mathfrak{W}_{11}:&\langle\mathrm{e}^{-nt}\partial_t+\mathrm{e}^{-nt}[n+\frac{\alpha n(n-1)}2\mathrm{e}^{-x}]\partial_x+\frac{n(n-1)}2\mathrm{e}^{-nt-x}\partial_u\rangle,
\end{eqnarray*}
where $n\in\mathbb{Z}$, $\alpha=0,\pm1$, $\gamma=\pm1$,
$\mathrm{sgn}(\cdot)$ is the standard sign function, the symbol $\tilde{\phi}(u)$ stands for either $u$
or an arbitrary real constant $c$, and
\begin{equation*}
\begin{split}
&f(x,u)=\mathrm{e}^x[4\mathrm{e}^{4x}-10\mathrm{e}^{3x}\tilde{\phi}-36\gamma\mathrm{e}^{2x}+2\mathrm{e}^{x}(31\gamma\tilde{\phi}+6\tilde{\phi}^3\pm6(4\gamma+\tilde{\phi}^2)^{3/2})\\[2mm]
&\qquad -64\gamma^2-54\gamma\tilde{\phi}^2-9\tilde{\phi}^4\mp9\tilde{\phi}(4\gamma+\tilde{\phi}^2)^{3/2}]r^{-1} \\[2mm]
&g(x,u)=\mathrm{e}^x[8\mathrm{e}^{4x}-16\mathrm{e}^{3x}\tilde{\phi}-2\mathrm{e}^{2x}(44\gamma+5\tilde{\phi}^2)+2\mathrm{e}^{x}(44\gamma\tilde{\phi}+9\tilde{\phi}^3\pm9(4\gamma+\tilde{\phi}^2)^{3/2})\\[2mm]
&\qquad -64\gamma^2-54\gamma\tilde{\phi}^2-9\tilde{\phi}^4\mp9\tilde{\phi}(4\gamma+\tilde{\phi}^2)^{3/2}]r^{-1},\\[2mm]
&r=4\mathrm{e}^{5x}-10\mathrm{e}^{4x}\tilde{\phi}-40\gamma\mathrm{e}^{3x}+10\mathrm{e}^{2x}(6\gamma\tilde{\phi}+\tilde{\phi}^3\pm(4\gamma+\tilde{\phi}^2)^{3/2})-
10\mathrm{e}^x(6\gamma^2+6\gamma\tilde{\phi}^2\\[2mm]
&\qquad+\tilde{\phi}^4\pm\tilde{\phi}(4\gamma+\tilde{\phi}^2)^{3/2})+30\gamma^2\tilde{\phi}
+20\gamma\tilde{\phi}^3+3\tilde{\phi}^5\pm(2\gamma+3\tilde{\phi}^2)(4\gamma+\tilde{\phi}^2)^{3/2}.
\end{split}
\end{equation*}

\end{theorem}

\proof To prove the theorem, it suffices to analyze all possible extensions
of the algebras \eqref{rep1} and \eqref{rep2}.

{\bf Case 1.}  Given the algebra \eqref{rep1} we make use of \eqref{5d} thus getting
\begin{equation*}
L_2=\mathrm{e}^{-2t}\partial_t,\qquad
L_{-2}=\mathrm{e}^{2t}\partial_t.
\end{equation*}
The remaining basis elements of the corresponding Witt algebra are easily obtained
through recursion, which yields $L_n=\mathrm{e}^{-nt}\partial_t,\
n\in\mathbb{Z}$. We arrive at the realization $\mathfrak{W}_1$ of
Theorem \ref{witt}.

{\bf Case 2.} Turn now to realization \eqref{rep2}. Inserting
$L_0,L_1,L_{-1}$ into the commutation relations
$[L_0,L_{-2}]=2L_{-2}$ and $[L_1,L_{-2}]=3L_{-1}$ and solving the
obtained PDEs, we have
\begin{equation*}
\begin{split}
   & L_{-2}=\mathrm{e}^{2t}(1+3\alpha\mathrm{e}^{-2x}+\psi_1(u)\mathrm{e}^{-3x})\partial_t
    +\mathrm{e}^{2t}(-2+3\phi(u)\mathrm{e}^{-x}+\psi_2(u)\mathrm{e}^{-2x}\\[2mm]
   &\quad\quad\quad  +\psi_1(u)\mathrm{e}^{-3x})\partial_x+\mathrm{e}^{2t}(3\beta\mathrm{e}^{-x}+
   \psi_3(u)\mathrm{e}^{-2x})\partial_u,
\end{split}
\end{equation*}
where $\psi_1,\psi_2,\psi_3$ are arbitrary smooth functions of $u$.

Using the relations $[L_0,L_2]=-2L_2$ and $[L_{-1},L_2]=-3L_1$ in a similar
fashion, we derive that
\begin{equation*}
   L_2=\mathrm{e}^{-2t}f(x,u)\partial_t+\mathrm{e}^{-2t}g(x,u)\partial_x+\mathrm{e}^{-2t}h(x,u)\partial_u,
\end{equation*}
$f,g,h$ satisfying the following system of PDEs
\begin{subequations}\label{eq1}
\begin{align}
& -3(\alpha\mathrm{e}^{-2x}+1)f+2\alpha\mathrm{e}^{-2x}g+(\phi\mathrm{e}^{-x}+\alpha\mathrm{e}^{-2x}-1)f_x+\beta\mathrm{e}^{-x}f_u+3=0,\label{1.1}\\[2mm]
&    (1-\phi\mathrm{e}^{-x}-\alpha\mathrm{e}^{-2x})f+(\phi\mathrm{e}^{-x}-2)g-\phi_u\mathrm{e}^{-x}h+(\phi\mathrm{e}^{-x}+\alpha\mathrm{e}^{-2x})g_x\label{1.2}\\[2mm]
&    \qquad +\beta\mathrm{e}^{-x}g_u+3=0,\nonumber\\[2mm]
 &   \beta\mathrm{e}^{-x}f-\beta\mathrm{e}^{-x}g+2(1+\alpha\mathrm{e}^{-2x})h-(\phi\mathrm{e}^{-x}+\alpha\mathrm{e}^{-2x}-1)h_x-\beta\mathrm{e}^{-x}h_u=0.\label{1.3}
\end{align}
\end{subequations}

Inserting the expressions for the basis elements $L_2$ and $L_{-2}$ into the
commutation relation $[L_2,L_{-2}]=4L_0$ yields three more PDEs
\begin{align}
 &4(\psi_1\mathrm{e}^{-3x}+3\alpha\mathrm{e}^{-2x}+1)f-3\mathrm{e}^{-2x}(\psi_1\mathrm{e}^{-x}+2\alpha)g+\mathrm{e}^{-3x}\dot{\psi_1}h\nonumber\\[2mm] &\quad -(\psi_1\mathrm{e}^{-3x}+\psi_2\mathrm{e}^{-2x}+3\phi\mathrm{e}^{-x}-2)f_x-\mathrm{e}^{-x}(\psi_3\mathrm{e}^{-x}+3\beta)f_u-4=0,\nonumber\\[2mm]
 &2(\psi_1\mathrm{e}^{-3x}+\psi_2\mathrm{e}^{-2x}+3\phi\mathrm{e}^{-x}-2)f-(\psi_1\mathrm{e}^{-3x}-2(3\alpha-\psi_2)\mathrm{e}^{-2x}+3\phi\mathrm{e}^{-x}-2)g\nonumber\\[2mm]
 &\quad+\mathrm{e}^{-x}(\dot{\psi_1}\mathrm{e}^{-2x}+\dot{\psi_2}\mathrm{e}^{-x}+3\dot{\phi})h-(\psi_1\mathrm{e}^{-3x}+\psi_2\mathrm{e}^{-2x}+3\phi\mathrm{e}^{-x}-2)g_x\label{eq2}\\[2mm]
 &\quad-\mathrm{e}^{-x}(\psi_3\mathrm{e}^{-x}+3\beta)g_u = 0,\nonumber\\[2mm]
 &2\mathrm{e}^{-x}(\psi_3\mathrm{e}^{-x}+3\beta)f-\mathrm{e}^{-x}(2\psi_3\mathrm{e}^{-x} +3\beta)g+(2\psi_1\mathrm{e}^{-3x} +(6\alpha+\dot{\psi_3})\mathrm{e}^{-2x}+2)h\nonumber\\[2mm]
 &\quad-(\psi_1\mathrm{e}^{-3x}+\psi_2\mathrm{e}^{-2x}+3\phi\mathrm{e}^{-x}-2)h_x-\mathrm{e}^{-x}(\psi_3\mathrm{e}^{-x}+3\beta)h_u = 0.\nonumber
\end{align}

To determine the forms of $L_2$ and $L_{-2}$, we have to solve Eqs.
\eqref{eq1} and \eqref{eq2}. It is straightforward to verify that
the relation
\begin{equation*}
  \Delta=\mathrm{e}^{-t-4x}[\beta\mathrm{e}^{3x} +
\psi_3\mathrm{e}^{2x}+(\beta\psi_2-\phi\psi_3-3\alpha\beta)\mathrm{e}^x
+\beta\psi_1-\alpha\psi_3]\neq0
\end{equation*}
is the necessary and sufficient condition for the system of
equations \eqref{eq1} and \eqref{eq2} to have the unique solution in
terms of $f_x$, $f_u$, $g_x$, $g_u$, $h_x$ and $h_u$. By this reason,
we need to differentiate between the cases $\Delta=0$ and $\Delta\neq0$.

{\bf Case 2.1.} Let $\Delta=0$ or, equivalently, $\beta=\psi_3=0$.
Eqs. \eqref{eq1} and \eqref{eq2} do not contain derivatives of the
functions $f,\ g,\ h$ with respect to $u$. That is why the
derivatives $f_x,\ g_x,\ h_x$ can be expressed in two different ways
using \eqref{eq1} and \eqref{eq2}. Equating the right-hand sides of
the two expressions for $h_x$ yields
\begin{equation*}
    h\mathrm{e}^x\frac{\mathrm{e}^{4x}-2\phi\mathrm{e}^{3x}-\psi_2\mathrm{e}^{2x}-2\psi_1\mathrm{e}^x+3\alpha^2+\phi\psi_1-\alpha\psi_2}
    {(\mathrm{e}^{2x}-\phi\mathrm{e}^x-\alpha)(2\mathrm{e}^{3x}-3\phi\mathrm{e}^{2x}-\psi_2\mathrm{e}^x-\psi_1)}=0.
\end{equation*}
Hence $h=0$. Similarly, the compatibility conditions for the
derivatives $f_x$ and $h_x$ give two more linear equations for the
functions $f$ and $g$. The determinant of the obtained system of
three linear equations does not vanish. Thus the system in question has
the unique solution for $f$ and $g$. Computing the derivatives of the
so obtained $f$ and $g$ with respect to $x$ and comparing the results
with the previously obtained expressions for $f_x$ and $g_x$, we arrive
at the equations
\begin{equation}
\label{eq10}
    (\psi_2-6\alpha)(\phi^3+\phi\psi_2+2\psi_1)\mathrm{e}^{11x}+F_{10}[x,u]=0,
\end{equation}
and
\begin{eqnarray}
\label{eq11}
    &&(10\phi^3\psi_1-3\alpha\phi^2(3\psi_2-8\alpha)+3\phi\psi_1(2\alpha+3\psi_2)+2(5\psi_1^2
    \\[2mm]
    &&\qquad -4\alpha(2\alpha^2-3\alpha_2+\psi_2^2)))\mathrm{e}^{10x}+F_9[x,u]=0.\nonumber
\end{eqnarray}
Hereafter $F_n[x,u]\ (n\in\mathbb{N})$ denotes a polynomial in
$\exp(x)$ of the power less than or equal to $n$. To find $f$ and
$g$ we need to construct the most general $\phi$ and
$\psi_i\ (i=1,2,3)$ satisfying Eqs. \eqref{eq10} and \eqref{eq11}.
If \eqref{eq10} holds, then at least one of the following equations
$\psi_2=6\alpha$ and $\psi_1=-(\phi^3+\phi\psi_2)/2$ should be
satisfied.

{\bf Case 2.1.1.} When $\psi_2=6\alpha$, Eqs. \eqref{eq10} and
\eqref{eq11} hold if and only if
\begin{equation*}
16\alpha^3+3\alpha^2\phi^2-6\alpha\phi\psi_1-\phi^3\psi_1-\psi_1^2=0,
\end{equation*}
whence $\psi_1=(-6\alpha\phi-\phi^3\pm(4\alpha+\phi^2)^\frac32)/2$.

{\bf Case 2.1.1.1.} Suppose now that $\psi_1=(-6\alpha\phi-\phi^3 -
(4\alpha+\phi^2)^\frac32))/2$. Provided $\alpha=0$, we have either
$\psi_1=0$ or $\psi_1=-\phi^3$. The case $\alpha=\psi_1=0$ leads to
$L_{-1}=\mathrm{e}^{t}\partial_t+\mathrm{e}^{t}(-1+\mathrm{e}^{-x}\phi)\partial_x$.
Making the equivalence transformation $\tilde{x}=x+X(u)$, we can
reduce $\phi$ to one of the forms $a=0,\pm1$. Thus
\begin{equation*}
f=1,\qquad g=2+a\mathrm{e}^{-x}.
\end{equation*}
Making use of the recurrence relations of the Witt algebra, we arrive at
the realization $\mathfrak{W}_2$.

Provided $\alpha=0$ and $\psi_1=-\phi^3$, we can reduce the function
$\phi$ to the form $b=0,\pm1$ with the equivalence transformation
$\tilde{x}=x+X(u)$. The case $b\not= 0$ gives rise to the following $f$ and $g$:
\begin{equation*}
   f=\frac{\mathrm{e}^x(\mathrm{e}^{2x}-3b\mathrm{e}^x+3b^2)}{(\mathrm{e}^x-b)^3},\qquad
   g=\frac{ \mathrm{e}^x(2\mathrm{e}^{x}-3b)}{(\mathrm{e}^x-b)^2}.
\end{equation*}
Hence the realization $\mathfrak{W}_3$ is obtained. Note that the case $b=0$ leads to
the particular case of $\mathfrak{W}_2$.

Assuming $\alpha=\pm1$, we have $\psi_1=(-6\alpha\phi-\phi^3-
(4\alpha+\phi^2)^\frac32)/2$ which yields $\mathfrak{W}_4$.

{\bf Case 2.1.1.2.} Let $\psi_1=(-6\alpha\phi-\phi^3 +(4\alpha +\phi^2)^\frac32))/2$.
If $\alpha=0$, then we have either $\psi_1=0$ or $\psi_1=-\phi^3$. This case has
already been considered when we analyzed the Case 2.1.1.1. When $\alpha=\pm1$,
we get the realization $\mathfrak{W}_4$.

{\bf Case 2.1.2.} If $\psi_1=-(\phi^3+\phi\psi_2)/2$, then Eq. \eqref{eq10}
takes the form
\begin{equation*}
(4\alpha+\phi^2)(\psi_2-(4\alpha-5\phi^2)/4)(\psi_2-(2\alpha-\phi^2))\mathrm{e}^{10x}+F_9[x,u]=0.
\end{equation*}
To solve the above equation, we need to consider the following three
subcases.

{\bf Case 2.1.2.1.} Given $\psi_2=(4\alpha-5\phi^2)/4$, Eqs.
\eqref{eq10} and \eqref{eq11} hold if and only if
\begin{equation*}
4\alpha+\phi^2=0.
\end{equation*}
Consequently $\alpha\leq0$ and $\phi=2b(-\alpha)^\frac12$ with
$b=\pm1$.

If $\alpha=-1$, we have $\phi=2b,\psi_1=2b,\psi_2=-6$ and furthermore
\begin{equation*}
f=\frac{\mathrm{e}^x(\mathrm{e}^x-2b)}{(\mathrm{e}^x-b)^2},\qquad g=\frac{2\mathrm{e}^x}{\mathrm{e}^x-b},
\end{equation*}
which leads to $\mathfrak{W}_5$.

In the case when $\alpha=0$ and $\phi=\psi_1=\psi_2=0$, we arrive at the
realization $\mathfrak{W}_2$ with $\alpha=0$.

{\bf Case 2.1.2.2.} Let $\psi_2=2\alpha-\phi^2$ and suppose that Eqs.
\eqref{eq10} and \eqref{eq11} hold. Provided $\alpha=0$ we can
transform $\phi$ to $b=\pm1$ (note that the case $b=0$ has
already been considered). Consequently,
\begin{equation*}
    f=1,\quad g=\frac{2\mathrm{e}^x-b}{\mathrm{e}^x-b}
\end{equation*}
and the realization $\mathfrak{W}_6$ is obtained.

Given $\alpha=\pm1$, we have
\begin{equation*}
    f=\frac{\mathrm{e}^x(\mathrm{e}^x-\phi)}{\mathrm{e}^{2x}-\mathrm{e}^x\phi-b},\qquad
    g=\frac{\mathrm{e}^x(2\mathrm{e}^x-\phi)}{\mathrm{e}^{2x}-\mathrm{e}^x\phi-b},
\end{equation*}
where $b=\pm1$. Since $\phi$ can be reduced to the form $\tilde{u}$
by the equivalence transformation $\tilde{u}=\phi$ with
$\dot{\phi}\neq0$, we get the realization $\mathfrak{W}_7$.

{\bf Case 2.1.2.3.} If $4\alpha+\phi^2=0$ and Eqs. \eqref{eq10} and
\eqref{eq11} holds, we get $\alpha\leq0$, whence $\alpha=0,-1$.

Given the relation $\alpha=0$, we can reduce $\phi$ to the form
$a=0,\pm1$. With this we obtain $f=1$ and $g=2-a\mathrm{e}^{-x}$,
thus getting $\mathfrak{W}_8$.

In the case when $\alpha=-1$, we have
\begin{equation*}
      f=\frac{\mathrm{e}^x(\mathrm{e}^{3x}-4\mathrm{e}^{2x}+5\mathrm{e}^x+4+\psi_2)}{(\mathrm{e}^x-1)^4},\quad
      g=\frac{\mathrm{e}^x(2\mathrm{e}^{2x}-4\mathrm{e}^x-4-\psi_2)}{(\mathrm{e}^x-1)^3}.
\end{equation*}
And what is more the function $\psi_2$ is reduced to the form
$\tilde{u}$ by the equivalence transformation $\tilde{u}=\psi_2$,
provided $\psi_2$ is a nonconstant function. As a result, we get $\mathfrak{W}_9$.

Summing up we conclude that the case $\Delta=0$ leads to the
realizations $\mathfrak{W}_i,\ i=2,3,\cdots,9$.

{\bf Case 2.2.} If $\Delta\neq0$, or equivalently,
$\beta^2+\psi_3^2\neq0$, then we can solve Eqs. \eqref{eq1}
and \eqref{eq2} for $f_x$, $f_u$, $g_x$, $g_u$, $h_x$ and $h_u$.
The compatibility conditions
\begin{equation*}
   f_{xu}-f_{ux}=0,\quad g_{xu}-g_{ux}=0,\quad h_{xu}-h_{ux}=0
\end{equation*}
can be rewritten as the following system of three linear equations
for the functions $f, g, h$
\begin{equation*}
\begin{split}
 &    a_1f+a_2g+a_3h+d_1=0,\\[2mm]
 &    b_1f+b_2g+b_3h+d_2=0,\\[2mm]
 &    c_1f+c_2g+c_3h+d_3=0.
\end{split}
\end{equation*}
Here $a_i,\ b_i,\ c_i,\ d_i,\ (i=1,2,3)$ are functions of $t,\ x,\
\phi,\ \psi_1,\ \psi_2,\ \psi_3$.

It is straightforward to verify that the above system has the unique
solution $f,\ g,\ h$ when $\beta^2+\psi_3^2\neq0$. We do not present
here the explicit formulae for these functions as they are very
cumbersome. Inserting these $f,\ g,\ h$ into Eq. \eqref{1.1} yields
\begin{equation*}
    \alpha\beta^6\mathrm{e}^{42x}+F_{41}[x,u]=0.
\end{equation*}
Consequently, we have either $\alpha=0$ or $\beta=0$.

{\bf Case 2.2.1.} If $\beta=0$, then Eq. \eqref{1.1} takes the form
\begin{equation*}
    \alpha\psi_3^6\mathrm{e}^{36x}+F_{35}[x,u]=0,
\end{equation*}
which gives $\alpha=0$ and $\psi_3\neq0$ (since $\Delta=0$ otherwise).
In view of these relations we can rewrite Eq. \eqref{eq2} as
follows
\begin{equation*}
\begin{split}
&   \psi_1\psi_3^6\mathrm{e}^{36x}+F_{35}[x,u]=0,\\[2mm]
&  (15\phi^2+2\psi_2)\psi_3^6\mathrm{e}^{37x}+F_{36}[x,u]=0,\\[2mm]
&   (57\phi^2-2\psi_2)\psi_3^7\mathrm{e}^{35x}+F_{34}[x,u]=0.
\end{split}
\end{equation*}
Hence we conclude that $\phi=\psi_1=\psi_2=0$. Inserting these
formulas into the initial Eqs. \eqref{eq1} and \eqref{eq2} and
solving the obtained system yield
\begin{equation*}
    f=1,\ g=2,\ h=-\mathrm{e}^{-2x}\psi_3.
\end{equation*}
The function $\psi_3$ can be reduced to the form $-1$ by the
transformation $\tilde{u}=U(u)$, where $\dot{U}=-1/\psi_3$.
As a result, we have $\mathfrak{W}_{10}$.

{\bf Case 2.2.2.} Provided $\alpha=0$, Eq. \eqref{1.3} turns into
\begin{equation*}
    \beta^5(4\beta\phi\psi_3-6\psi_3^2+\beta^2\dot{\psi_3})
    \mathrm{e}^{41x}+30\beta^5\phi\psi_3^2\mathrm{e}^{40x}+F_{39}[x,u]=0.
\end{equation*}
Note that the case $\alpha=\beta=0$ has already been analyzed in Case 2.2.1.
Consequently, without any loss of generality we can restrict our
considerations to the two cases $\psi_3=0$, $\beta=1$ and $\phi=0$,
$\beta=1$.

If $\psi_3=0$, then it follows from \eqref{eq2} and \eqref{1.3} that
$\psi_1=\psi_2=0$. In view of these relations, we get from \eqref{eq1} and
\eqref{eq2} that
\begin{equation*}
    f=1,\ g=2+\mathrm{e}^{-x}\phi,\ h=\mathrm{e}^{-x}.
\end{equation*}
What is more, the function $\phi$ can be reduced to one the forms
$0,\pm1$ by the equivalence transformations $\tilde{x}=x+X(u)$ and
$\tilde{u}=U(u)$. Hence we get the realization $\mathfrak{W}_{11}$.

In the case $\phi=0$, Eqs. \eqref{eq1} and \eqref{eq2} are
incompatible.

We check by direct computation that the realizations $\mathfrak{W}_i\
(i=1,2,\cdots,11)$ cannot be mapped one into another by a
transformation of the form \eqref{tr}. Consequently, they are inequivalent.
This completes the proof of the theorem.\qed

While proving Theorem \ref{witt}, we have also obtained the exhaustive description
of the Witt algebras in the spaces $\mathbb{R}^1$ and $\mathbb{R}^2$, as a
by-product.
\begin{theorem}\label{witt_r1}
There is only one inequivalent realization, $\mathfrak{W}_1$, of the Witt algebra
in the space $\mathbb{R}$.
\end{theorem}
\begin{theorem}\label{witt_r2}
The realizations $\mathfrak{W}_1$--$\mathfrak{W}_9$ with $\tilde{\phi}=c\in \mathbb{R}$ exhaust
the list of inequivalent realizations of Witt algebra in the space $\mathbb{R}^2$.
\end{theorem}

\section{Realizations of the Virasoro algebra}

To construct all inequivalent realizations of the Virasoro algebra
$\mathfrak{V}$, we need to extend inequivalent Witt algebras in
Theorem \ref{witt} by all possible nonzero central elements $C$.
In this section, we will prove that there are no realizations of
the Virasoro algebra with nonzero central element in the
space $\mathbb{R}^3$.

Let us begin by constructing all possible central extensions of the
subalgebra $\langle L_0, L_1, L_{-1} \rangle$. According to Lemma \ref{3d},
it suffices to consider the algebras \eqref{rep1} and \eqref{rep2}.

{\bf Case 1.} Given the realization \eqref{rep1}, we have
\begin{equation*}
  L_0=\partial_t,\ L_1=\mathrm{e}^{-t}\partial_t,\
  L_{-1}=\mathrm{e}^t\partial_t.
\end{equation*}
Letting the basis element $C$ be of the general form \eqref{basis} and inserting
it into the commutation relations $[L_i,C]=0,\
(i=0,1,-1)$ yield
\begin{equation*}
    C=\xi(x,u)\partial_x+\eta(x,u)\partial_u,\quad \xi^2+\eta^2\neq0.
\end{equation*}
Applying the transformation
\begin{equation*}
    \tilde{t}=t,\quad \tilde{x}=X(x,u),\quad \tilde{u}=U(x,u),
\end{equation*}
which preserves $L_0$, $L_1$ and $L_{-1}$, to the
central element $C$, we get
\begin{equation*}
    C\to\widetilde{C}=(\xi X_x+\eta X_u)\partial_{\tilde{x}}+(\xi U_x+\eta U_u)\partial_{\tilde{u}}.
\end{equation*}
We choose solutions of the equations
\begin{equation*}
    \xi X_x+\eta X_u=0,\quad\xi U_x+\eta U_u=1
\end{equation*}
as $X$ and $U$, and get $C=\partial_u$.

Proceed now to constructing $L_2$. Making use of the commutation
relations $[L_0,L_2]=-2L_2$, $[L_{-1},L_2]=-3L_1$ and $[L_2,C]=0$,
yields $L_2=\mathrm{e}^{-2t}\partial_t$. Next, let $L_{-2}$ be of the
form \eqref{basis}. With this $L_{-2}$, the commutation
relations \eqref{5d} involving $L_{-2}$ are equivalent to
an over-determined system of PDEs for the unknown functions $\tau$,
$\xi$ and $\eta$. This system turns out to be incompatible.
Hence realization \eqref{rep1} cannot be extended up to a realization
of the Virasoro algebra with nonzero central element.

{\bf Case 2.} Consider now the algebra \eqref{rep2}. Since $C$ should
commute with $L_0$ and $L_1$, we have
\begin{equation*}
C=f(u)\mathrm{e}^{-x}\partial_t+(g(u)+f(u)\mathrm{e}^{-x})\partial_x+h(u)\partial_u,
\end{equation*}
where $f$, $g$ and $h$ are arbitrary smooth functions.
Acting by transformation \eqref{tr2}, that does not alter $L_0$ and $L_1$,
on $C$ gives
 \begin{equation*}
    \widetilde{C}=f(u)\mathrm{e}^{-x}\partial_{\tilde{t}}+(g(u)+f(u)\mathrm{e}^{-x}+h(u)\dot{X}(u))\partial_{\tilde{x}}
    +h(u)\dot{U}(u)\partial_{\tilde{u}}.
\end{equation*}
To further simplify $\widetilde{C}$, we analyze the cases $f(u)\neq0$
and $f(u)=0$ separately.

If $f(u)\neq0$, then choosing $X(u)=-\ln{|f(u)|}$ we have
$\widetilde{C}=\mathrm{e}^{-\tilde{x}}\partial_{\tilde{t}}+
(\mathrm{e}^{-\tilde{x}}+\beta(g+h\dot{X}))\partial_{\tilde{x}}
+\beta h\dot{U}\partial_{\tilde{u}}$, where $\beta=\pm1$. Provided
$h=0$ and $\dot{g}\neq0$, we can make the transformation $\tilde{u}=g(u)$
and thus get $C_1=\mathrm{e}^{-x}\partial_t +(\mathrm{e}^{-x}+
u)\partial_x$. The case $h=\dot{g}=0$ leads to
$C_2=\mathrm{e}^{-x}\partial_t + (\mathrm{e}^{-x}
+\lambda)\partial_x$, where $\lambda$ is an arbitrary constant.
Next, if $h\neq0$ then we choose solutions of the equations
$g+h\dot{X}=0$ and $h\dot{U}=1/\beta$ as $X$ and $U$ and thus
$C_3=\mathrm{e}^{-x}\partial_t + \mathrm{e}^{-x}\partial_x +
\partial_u$ is obtained.

Provided $f(u)=0$, we have $\widetilde{C}=(g +
h\dot{X})\partial_{\tilde{x}} + h\dot{U}\partial_{\tilde{u}}$.
If $h\neq0$, we can reduce $C_4$ to the form $\partial_u$ by
a suitable choice of $X$ and $U$.

Given the condition $h=0$, we have $\widetilde{C}=g\partial_{\tilde{x}}$.
If $g$ is not a constant, then selecting $U=g(u)$ yields
$C_5=u\partial_x$. The case of constant $g$ leads to
$C_6=\partial_x$.

Summing up, we conclude that there exist six inequivalent nonzero
central element $C$ for the case when
$L_0=\partial_t$ and $L_1=\mathrm{e}^{-t}\partial_t +
\mathrm{e}^{-t}\partial_x$. Now we need to extend the realizations
$\langle L_0,L_1,C_i \rangle,\ (i=1,2,\cdots,6)$ up to realizations of
the full Virasoro algebra. Here we present the calculation details for
the case $i=1$ only. The remaining five cases are handled in a similar
fashion.

To extend $\langle L_0,L_1,C_1 \rangle$ up to a realization of the full Virasoro
algebra, we need to construct all possible realizations of $L_{-1}$. Taking into account
\eqref{sub} we have
\begin{equation*}
    L_{-1}=\frac{\mathrm{e}^{t-2x}(u^2\mathrm{e}^{2x}-1)}{u^2}\partial_t-
    \frac{\mathrm{e}^{t-2x}(u\mathrm{e}^{x}+1)^2}{u^2}\partial_x.
\end{equation*}
With $L_{-1}$ in hand, we proceed to constructing $L_2$. Using the commutation
relations \eqref{5d} yields
\begin{equation*}
    L_2=\frac{u\mathrm{e}^x(u\mathrm{e}^x+2)}{\mathrm{e}^{2t}(u\mathrm{e}^x+1)^2}\partial_t+
    \frac{2u\mathrm{e}^x}{\mathrm{e}^{2t}(u\mathrm{e}^x+1)}\partial_x.
\end{equation*}
While constructing $L_{-2}$, we arrive at the incompatible system of PDEs for
its coefficients. Hence, the algebra $\langle L_0,L_1,C_1 \rangle$
cannot be extended to a realization of the full Virasoro algebra. The same result
holds for the remaining realizations $C_2,C_3,\ldots,C_6$.
\begin{theorem}
There are no realizations of the Virasoro algebra with nonzero central
element $C$ in the space $\mathbb{R}^n,\ (n=1,2,3)$.
\end{theorem}

\section{PDEs invariant under the Witt algebras}

In this section we construct a number of new classes of second-order evolution equations
in $\mathbb{R}^2$ that admit the Witt algebra. Given a
realization of the Witt algebra, we can apply the Lie infinitesimal approach
to construct the corresponding invariant equation \cite{olv86,ovs82}. Differential
equation
\begin{equation*}
    F(t,x,u,u_t,u_x,u_{tt},u_{tx},u_{xx})=0
\end{equation*}
is invariant with respect to the Witt algebra $\langle L_n\rangle$ if and only
if the condition
\begin{equation*}
    \text{pr}^{(2)}L_n(F)|_{F=0}=0
\end{equation*}
holds for any $n\in\mathbb{N}$, where $\text{pr}^{(2)}L_n$ is the
second-order prolongation of the vector field $L_n$, that is
\begin{equation*}
\text{pr}^{(2)}L_n=L_n+\eta^t\partial_{u_t}+\eta^x\partial_{u_x}+\eta^{tt}\partial_{u_{tt}}+
\eta^{tx}\partial_{u_{tx}}+\eta^{xx}\partial_{u_{xx}}
\end{equation*}
with
\begin{align*}
        \eta^t      &= D_t(\eta)-u_tD_t(\tau)-u_xD_t(\xi), \\[2mm]
        \eta^x      &= D_x(\eta)-u_tD_x(\tau)-u_xD_x(\xi), \\[2mm]
        \eta^{tt}   &= D_t(\eta^t)-u_{tt}D_t(\tau)-u_{tx}D_t(\xi), \\[2mm]
        \eta^{tx}   &= D_x(\eta^t)-u_{tt}D_x(\tau)-u_{tx}D_x(\xi), \\[2mm]
        \eta^{xx}   &= D_x(\eta^x)-u_{xt}D_x(\tau)-u_{xx}D_x(\xi).
\end{align*}
Here the symbols $D_t$ and $D_x$ stand for the total differentiation
operators with respect to $t$ and $x$, correspondingly,
\begin{align*}
  D_t &= \partial_t+u_t\partial_u+u_{tt}\partial_{u_t}+u_{xt}\partial_{u_x}+\cdots,
  \\[2mm]
  D_x &= \partial_x+u_x\partial_u+u_{tx}\partial_{u_t}+u_{xx}\partial_{u_x}+\cdots.
\end{align*}

As an example, we present the procedure of constructing
$\mathfrak{W}_1$ invariant equations in detail. Utilizing the formulas above, we obtain
\begin{equation}\label{Q_2}
    \text{pr}^{(2)}L_n=\mathrm{e}^{-nt}\partial_t+n\mathrm{e}^{-nt}u_t\partial_{u_t}
    +(2n\mathrm{e}^{-nt}u_{tt}-n^2\mathrm{e}^{-nt}u_t)\partial_{u_{tt}}
    +n\mathrm{e}^{-nt}u_{tx}\partial_{u_{tx}}.
\end{equation}
The next step is computing the full set of functionally-independent
second-order differential invariants,
$I_m(t,x,u,u_t,u_x,u_{tt},u_{tx},u_{xx})\ (m=1,2,\cdots,7),$
associated with $L_n$. To get $I_m$, we need to solve the corresponding
characteristic equations
\begin{equation*}
    \frac{\mathrm{d}t}{\mathrm{e}^{-nt}}=\frac{\mathrm{d}x}{0}=
    \frac{\mathrm{d}u}{0}=\frac{\mathrm{d}u_t}{n\mathrm{e}^{-nt}u_t}
    =\frac{\mathrm{d}u_x}{0}=\frac{\mathrm{d}u_{tt}}
    {2n\mathrm{e}^{-nt}u_{tt}-n^2\mathrm{e}^{-nt}u_t}
    =\frac{\mathrm{d}u_{tx}}{n\mathrm{e}^{-nt}u_{tx}}
    =\frac{\mathrm{d}u_{xx}}{0}.
\end{equation*}
Integration of the above equations yields
\begin{equation*}
  I_1=x,\ I_2=u,\ I_3=u_x,\ I_4=u_{xx},\ I_5=\frac{u_{tx}}{u_t},\
  I_6=\mathrm{e}^{-nt} u_t,\ I_7=\mathrm{e}^{-2nt}u_{tt}-n\mathrm{e}^{-2nt}u_t.
\end{equation*}
Hence the most general $L_n$-invariant equation is of the form
\begin{equation*}
   F(I_1,I_2,\cdots,I_7)=0.
\end{equation*}
Since this equation should be invariant under every basis element
of the Witt algebra $\mathfrak{W}_1$, it must be
independent of $n$. To meet this requirement, function $F$ has to
be independent of $I_6$ and $I_7$. Thus the most general
second-order PDE invariant under $\mathfrak{W}_1$ has the form
\begin{equation*}
   F(I_1,I_2,I_3,I_4,I_5)=0,
\end{equation*}
or, equivalently,
\begin{equation*}
F\left(x,u,u_x,u_{xx},\frac{u_{tx}}{u_t}\right)=0.
\end{equation*}
What is more, we have succeeded in constructing the general forms of PDEs
invariant under $\mathfrak{W}_2$, $\mathfrak{W}_6$, $\mathfrak{W}_8$ and
$\mathfrak{W}_{10}$. We list the corresponding invariant equations in
Table 1, where $F$ is an arbitrary smooth real-valued function.
\begin{center}{{\bf Table 1.} Second-order PDEs admitting Witt algebra}\end{center}
\begin{center}
\begin{tabular}{cl}\hline
Symmetry algebra           &\qquad\qquad Invariant equation\\[.10cm]\hline
$\mathfrak{W}_1$           &$F(x,u,u_x,u_{xx},\frac{u_{tx}}{u_t})=0$\\[2mm]
$\mathfrak{W}_2$                      &\parbox{10cm}{$F(u,u_x,u_{xx},\frac{u_tu_{xx}-u_xu_{tx}}{\mathrm{e}^xu_x})=0,\quad\alpha=0$\\
                                   $F(u,\frac{u_{xx}-u_x}{u_x^2},\frac{u_tu_x-u_tu_{xx}+u_xu_{tx}+u_x^2}{\mathrm{e}^xu_x}-2\alpha u_x)=0,\ \ \alpha=\pm1$}     \\[2mm]
$\mathfrak{W}_6$           &$F(u,\frac{\gamma(u_{xx}+u_{tx})-\mathrm{e}^x(u_x+u_{xx})}{u_x(\gamma(u_t+u_x)-\mathrm{e}^xu_x)})$\\[2mm]                     $\mathfrak{W}_8$                      &$F(u,\frac{u_{xx}-2u_x}{u_x^2})=0$\\[2mm]
$\mathfrak{W}_{10}$                   &$F(u_x+2u,u_{xx}-4u)=0$\\[0.10cm]\hline
\end{tabular}
\end{center}

\section{The direct sums of the Witt algebras}
This section is devoted to classification of realizations of the direct sum
of the Witt algebras in $\mathbb{R}^3$. We obtain the complete description of
inequivalent realizations of the direct sums of two Witt algebras.

According to Theorem \ref{witt}, it suffices to consider realizations
of the form
\begin{equation*}
    \mathfrak{W}_i\oplus\langle \tilde{L}_n,\ n\in\mathbb{Z}\rangle,\qquad i=1,2,\cdots,11,
\end{equation*}
where $\mathfrak{W}_i$ are given in Theorem \ref{witt} and $\tilde{L}_n$,\ $n\in\mathbb{Z}$
are basis elements of the Witt algebra commuting with the corresponding
realization $\mathfrak{W}_i$.

We begin by considering the realization $\mathfrak{W}_1\oplus\langle \tilde{L}_n\rangle$.
Let us choose $\tilde{L}_n$ in the general form \eqref{basis}. As $\tilde{L}_n$ should commute
with $\mathfrak{W}_1$, we have
\begin{equation}\label{w1_Q}
    \tilde{L}_n=f_n(x,u)\partial_x+g_n(x,u)\partial_u.
\end{equation}
Here $f_n$ and $g_n$ are arbitrary smooth functions. We have established in
Section 3 that the realizations $\langle \tilde{L}_n\rangle$ with basis
operators \eqref{w1_Q} exhaust the list of inequivalent realizations of the
Witt algebra in the space $\mathbb{R}^2$ of the variables $t$ and $x$.
Consequently, we can replace $t,x$ with $x,u$ respectively in
$\mathfrak{W}_i,\ (i=1,\cdots,9)$ presented in Theorem \ref{witt_r2},
thus getting all possible inequivalent realizations of
$\mathfrak{W}_1\oplus\langle \tilde{L}_n\rangle$ .

The realizations $\mathfrak{W}_i,\ (i=2,\cdots,11)$ are handled in the same way.
We skip rather tedious and cumbersome computations and present the final results
in the assertion below.
\begin{theorem}
Any realization of the direct sum of two Witt algebras in $\mathbb{R}^3$ is equivalent to one
of the realizations, $\{\mathfrak{D}_i,\ i=1,2,\cdots,10\}$, below
\begin{align*}
\mathfrak{D}_1:\qquad &      \langle\mathrm{e}^{-mt}\partial_t\rangle\oplus\langle\mathrm{e}^{-nx}\partial_x\rangle,\\[2mm]
\mathfrak{D}_2:\qquad & \langle\mathrm{e}^{-mt}\partial_t\rangle\oplus\langle\mathrm{e}^{-nx}\partial_x+n\mathrm{e}^{-nx}\partial_u\rangle,\\[2mm]
\mathfrak{D}_3:\qquad &       \langle\mathrm{e}^{-mt}\partial_t+m\mathrm{e}^{-mt}\partial_x\rangle\oplus\langle n\mathrm{e}^{-nu}\partial_x+\mathrm{e}^{-nu}\partial_u\rangle,\\[2mm]
\mathfrak{D}_4:\qquad &      \langle\mathrm{e}^{-mt}\partial_t\rangle\oplus\langle \mathrm{e}^{-nx}\partial_x+\gamma\mathrm{e}^{-nx}[\mathrm{e}^{nu}-(\mathrm{e}^u-\gamma)^n]
(\mathrm{e}^u-\gamma)^{1-n}\partial_u\rangle,\\[2mm]
\mathfrak{D}_5:\qquad &       \langle\mathrm{e}^{-mt}\partial_t\rangle\oplus\langle \mathrm{e}^{-nx}\partial_x+\mathrm{e}^{-nx}[n-\mathrm{sgn}(n)\frac{\gamma}{2}\sum_{j=1}^{|n|-1}j(j+1)
\mathrm{e}^{-2u}]\partial_u\rangle,\\[2mm]
\mathfrak{D}_6:\qquad &     \langle\mathrm{e}^{-mt}\partial_t\rangle\oplus\langle\mathrm{e}^{-nx+(n-1)u}(\mathrm{e}^u\pm n)(\mathrm{e}^u\pm1)^{-n}\partial_x\\[2mm]
&\quad +n\mathrm{e}^{-nx+(n-1)u}(\mathrm{e}^u\pm1)^{1-n}\partial_u\rangle,\\[2mm]
\mathfrak{D}_7:\qquad &     \langle\mathrm{e}^{-mt}\partial_t\rangle\oplus\langle\mathrm{e}^{-nx+(n-1)u}
[\mathrm{e}^{2u}-(n+1)\gamma\mathrm{e}^u+\frac12n(n+1)]
       (\mathrm{e}^u-\gamma)^{-n-1}\partial_x\\
  &\quad +\mathrm{e}^{-nx+(n-1)u}[n\mathrm{e}^u-\frac12n(n+1)\gamma](\mathrm{e}^u-\gamma)^{-n}\partial_u\rangle,\\[2mm]
\mathfrak{D}_8:\qquad &  \langle\mathrm{e}^{-mt}\partial_t\rangle\oplus\langle J_1\partial_x+J_2\partial_u\rangle,\\[2mm]
\mathfrak{D}_9:\qquad &  \langle\mathrm{e}^{-mt}\partial_t\rangle\oplus\widetilde{\mathfrak{W}}_4,\\[2mm]
\mathfrak{D}_{10}:\qquad &  \langle\mathrm{e}^{-mt}\partial_t\rangle\oplus\widetilde{\mathfrak{W}}_7.
\end{align*}
Here
\begin{align*}
    J_1&=\frac{\mathrm{e}^{-nx+(n-1)u}}{(\mathrm{e}^u-1)^{n+2}}[(-1+\sum_{j=1}^{|n|-1}(2j+1))n+(2n+1)\mathrm{e}^u
         -(n+2)\mathrm{e}^{2u}+\mathrm{e}^{3u}\\
         &\quad+\mathrm{sgn}(n)\frac{c}2\sum_{j=1}^{|n|-1}j(j+1)],\\
    J_2&=\frac{\mathrm{e}^{-nx+(n-1)u}}{(\mathrm{e}^u-1)^{n+1}}
   [(1-\sum_{j=1}^{|n|-1}(2j+1))n-2n\mathrm{e}^u+n\mathrm{e}^{2u}-\mathrm{sgn}(n)\frac{c}2\sum_{j=1}^{|n|-1}j(j+1)],
\end{align*}
$n\in\mathbb{Z}$, $m\in\mathbb{Z}$, $c\in\mathbb{R}$ and the symbols $\widetilde{\mathfrak{W}}_4$ and
$\widetilde{\mathfrak{W}}_7$ stand for the realizations obtained from $\mathfrak{W}_4$
and $\mathfrak{W}_7$ listed in Theorem \ref{witt_r2} by replacing $(t,x)$ with $(x,u)$.
\end{theorem}

Analysis of second-order differential equations invariant under the direct sum of
the Witt algebras yields that there are no equations that admit realizations
$\mathfrak{D}_4$, $\mathfrak{D}_5$ and $\mathfrak{D}_7$--$\mathfrak{D}_{10}$.
The remaining realizations of the direct sum of the Witt algebras gives rise to
the following invariant nonlinear PDEs:
\begin{eqnarray}
&\mathfrak{D}_1:&\quad F\left(u,\frac{u_{tx}}{u_tu_x}\right)=0,\label{d1}\\[2mm]
&\mathfrak{D}_2:&\quad F\left(\frac{u_{tx}}{u_t}\mathrm{e}^{-u}\right)=0,\label{d2}\\[2mm]
&\mathfrak{D}_3:&\quad F\left(\frac{u_tu_{xx}-u_xu_{tx}}{u_x^3}\mathrm{e}^{-x}\right)=0,\label{d3}\\[2mm]
&\mathfrak{D}_6:&\quad F\left(\frac{u_{tx}(1-u_x\pm\mathrm{e}^u)+u_t(u_{xx}-u_x^2+u_x)}{u_t(\mathrm{e}^{2u}+(u_x-1)
                              (u_x-1\mp2\mathrm{e}^u))}\right)=0.\label{d6}
\end{eqnarray}
Here $F$ is an arbitrary smooth real-valued function.

Let us reiterate, {\it any} second-order PDE, in two independent variables,
which is invariant under the direct sum of the Witt algebras, is equivalent to
one of the equations, (\ref{d1})--(\ref{d6}).

PDEs (\ref{d1})--(\ref{d6}) are classically integrable in the sense that they
admit infinite symmetry groups involving two arbitrary functions of one
variable.

Eq. (\ref{d1}) can be rewritten in the equivalent form
$$
u_{tx} = f(u)u_tu_x.
$$
Making the change of variables $u\to \tilde u = U(u)$ with
appropriately chosen $U(u)$ reduces the above PDE to the linear wave
equation $\tilde u_{tx}=0$.

Without any loss of generality, we can rewrite (\ref{d2}) in the form
$$
u_{tx} = \lambda {u_t}\mathrm{e}^{u},\quad \lambda \in {\mathbb R}.
$$
Integrating it above with respect to $t$ yields
$$
u_x = \lambda\mathrm{e}^{u}+\frac{g''(x)}{g'(x)},
$$
where $g(x)$ is an arbitrary smooth function satisfying $g'\neq0$. The obtained equation
can be represented in the equivalent form
$$
(u-\ln g'(x))_x=\lambda \mathrm{e}^{\left(u-\ln g'(x)\right)}\mathrm{e}^{\ln g'(x)}.
$$
It is straightforward to integrate the equation above and thus get
the general solution of the initial nonlinear PDE (\ref{d2})
$$
u(t,x)=\ln\frac{g'(x)}{h(t)-\lambda g(x)},
$$
where $g, h$ are arbitrary smooth real-valued functions with $g'\neq0$.

Eq. (\ref{d3}) is equivalent to the following PDE:
$$
u_tu_{xx}-u_xu_{tx}=\lambda\mathrm{e}^{x}u_x^3,\quad \lambda \in {\mathbb R}.
$$
The hodograph transformation $x\to u, u\to x$ and re-scaling $t\to\lambda t$
reduce it to the Liouville equation (\ref{lio}), which is known
to be integrable.

To the best of our knowledge, Eq. (\ref{d6}) is the new
classically integrable nonlinear PDE.

\section {Concluding Remarks}

In this paper, we perform the exhaustive classification of
the realizations of the Witt and Virasoro algebras by Lie
vector fields in the space $\mathbb{R}^n$ with $n=1,2,3$.
The complete lists of inequivalent realizations are given in
Theorems 1--5.

The main classification results can be briefly summarized as follows:
\begin{itemize}
\item There exists only one inequivalent realization of the Witt algebra
in $\mathbb{R}$.
\item There are nine inequivalent realizations of the Witt algebra in $\mathbb{R}^2$.
\item There exist eleven inequivalent realizations of the Witt algebra in
$\mathbb{R}^3$ space.
\item There are no realizations of the Virasoro algebra with nonzero central element
in the space $\mathbb{R}^n$ with $n\le 3$.
\item There exist ten inequivalent realizations of the direct sum of the Witt algebras
in $\mathbb{R}^3$.
\end{itemize}

As an application, we construct a number of new nonlinear PDEs which are invariant
under various realizations of the Witt algebra.

What is more, we completely classify the nonlinear second-order
PDEs in two independent variables admitting direct sums of the Witt algebras
and obtain four canonical invariant equations \eqref{d1}--\eqref{d6} which
possess infinite-dimensional algebras involving two arbitrary functions. As we
have mentioned before, the well-known massless wave and Liouville equations
are typical examples of such PDEs. Among them, Eqs. \eqref{d1}-\eqref{d3}
are well-known, while the nonlinear PDE \eqref{d6} is seemingly new.

Furthermore, since Virasoro algebra is a subalgebra of the Kac-Moody-Virasoro algebra,
the results obtained here can be directly applied to classify the integrable
KP type equations in $(1+2)$ dimensions. The starting point would be describing
inequivalent realizations of the Kac-Moody-Virasoro algebras by differential
operators in $\mathbb{R}^4$.

This problem is under study now and will be reported in our future publications.

\end{document}